\documentclass[reprint,amsmath,amssymb,aps]{revtex4-2}
\usepackage{graphicx}
\usepackage{dcolumn}
\usepackage{bm}
\usepackage{natbib}
\usepackage{color}
\usepackage{booktabs}
\usepackage{relsize}

\newcommand{\ra}{\rightarrow}

\newcommand{\lra}{\leftrightarrows}

\newcommand{\lap}{\mbox{$\cal L$}}

\newcommand{\pr}{\mbox{Pr}}

\renewcommand{\thetable}{\arabic{table}}
\newcommand{\beginsupplement}{%
        \setcounter{table}{0}
        \renewcommand{\thetable}{S\arabic{table}}%
        \setcounter{figure}{0}
        \renewcommand{\thefigure}{S\arabic{figure}}%
     }
\newcommand{\dkl}{\text{D}_{\textsubscript{KL}}}

\begin{document}

\preprint{APS/123-QED}

\title{Universal Bounds on Information-Processing Capabilities of Markov Processes }

\author{U\u{g}ur \c{C}etiner$^{1}$}  
\author{Jeremy Gunawardena$^{1}$}  

\affiliation{%
$^1$ Department of Systems Biology, Harvard Medical School, Boston, MA 02115, USA.
}%
\begin{abstract}
We consider a finite-state, continuous-time Markov process, represented in the ``linear framework" by a directed graph with labelled edges which specifies the infinitesimal generator of the process. If the graph is strongly connected, the process has a unique steady-state probability distribution, $p$, which may not be one of thermodynamic equilibrium. If the label (rate) of any edge (transition) is perturbed, to reach the new steady-state probability distribution $p'$, we find that the Kullback-Leibler (KL) divergence between these distributions is bounded by the change in the thermodynamic affinity, $\Delta A(C)$, of any cycle, $C$, that includes the altered transition, $\text{D\textsubscript{KL}}(p'||p) \leq |\Delta A(C)|$, irrespective of the structure of the graph. It follows that, if an equilibrium distribution is shifted away from equilibrium by perturbing a single rate, then the free energy difference between these distributions is similarly bounded $F^{neq}-F^{eq}\leq |\Delta A(C)|$. Our analysis reveals universal, energy-induced bounds on the information-processing capabilities of Markov systems operating arbitrarily far from thermodynamic equilibrium.
\end{abstract}

\maketitle
\section*{Introduction} 

Equilibrium thermodynamics is unmatched in its simplicity and applicability. It has been used to describe systems at various scales, ranging from black holes to bacterial gene regulation. However, at equilibrium, each reaction is balanced with a reverse one, a special condition known as detailed balance \citep{lewis1925new}. As a consequence, neither spatial nor temporal directionality can be achieved at equilibrium. Living systems must exchange energy or matter with their surroundings to break detailed balance and carry on the business of life \citep{schrodinger1944life}. Molecular motors, for example, convert the free energy of hydrolysis of ATP into mechanical work \citep{kolomeisky2007molecular}. The significance of energy expenditure in more abstract tasks, such as cellular information processing, remains more elusive, in part because information processing can be undertaken at thermodynamic equilibrium. John Hopfield’s pioneering work on discrimination by kinetic proofreading showed that certain cellular information-processing capabilities, such as error correction in the synthesis of biological macromolecules, can be improved by energy expenditure beyond their equilibrium limits \citep{hopfield1974kinetic}. However, the underlying physics behind this phenomenon is not yet comprehensively understood.

In previous work, we have developed a graph-theoretic approach to Markov processes \citep{gunawardena2012linear,mirzaev2013laplacian,kmg22}, in which steady-state probabilities can be expressed as rational functions of the transition rates. If the process reaches thermodynamic equilibrium, the steady-state formulas are equivalent to those of equilibrium statistical mechanics but, importantly, they remain valid arbitrarily far from equilibrium. This has provided the means to analyse non-equilibrium cellular information processing \citep{wjg19,kmg22} but progress in this direction has been hampered by the extraordinary algebraic complexity of non-equilibrium steady states and the difficulty of interpreting the algebra in thermodynamic terms. 

In recent work, we have developed a way to reorganise this complexity within the linear framework, so as to render it tractable and give it thermodynamic meaning \citep{ccetiner2022reformulating}. At thermodynamic equilibrium, the steady-state probability of a graph vertex, $i$, is proportional to the exponential of the \emph{path entropy} along any minimal path to $i$ from some arbitrary reference vertex, $1$. (We use a different sign convention in Eq.\ref{eql} below.) The independence of the path entropy from the chosen path is equivalent to detailed balance. We show that, away from equilibrium, the steady-state probability of vertex $i$ is an average of the exponential of the minimal path entropies from $1$ to $i$, where the average is taken over a probability distribution on the spanning trees of the graph rooted at $1$, which we call the \emph{arboreal distribution} (Eq.\ref{NessFinal1}). This reformulation smoothly generalises the equilibrium formula and gives it a thermodynamic interpretation. It has made it possible to undertake non-equilibrium calculations that were previously out of reach \citep{ccetiner2022reformulating}.
\vspace{-0.01cm}

Here, we exploit this reformulation to derive a set of information-theoretic inequalities. We show, in particular, that the Kullback-Leibler (KL) divergence between any steady state probability distribution, $p$, and any other state-state probability distribution, $p'$, that is obtained by perturbing a single edge label, is bounded by the change in thermodynamic affinity, $\Delta A(C)$ of any cycle, $C$, that contains that edge,  $\text{D\textsubscript{KL}}(p'||p) \leq |\Delta A(C)|$ (Eq.\ref{KLD}). KL divergence is an information-theoretic measure of the ``distance" between two probability distributions. It has had broad applications in studies of non-equilibrium systems, which we briefly review in the Discussion. This bound is universal, in the sense of holding for any strongly-connected graph, irrespective of its overall structure, and it is valid even if the underlying system operates far from thermodynamic equilibrium.

\section*{Background of previous work}

\subsection*{The linear framework}

\begin{figure}
\includegraphics[width=0.45\textwidth]{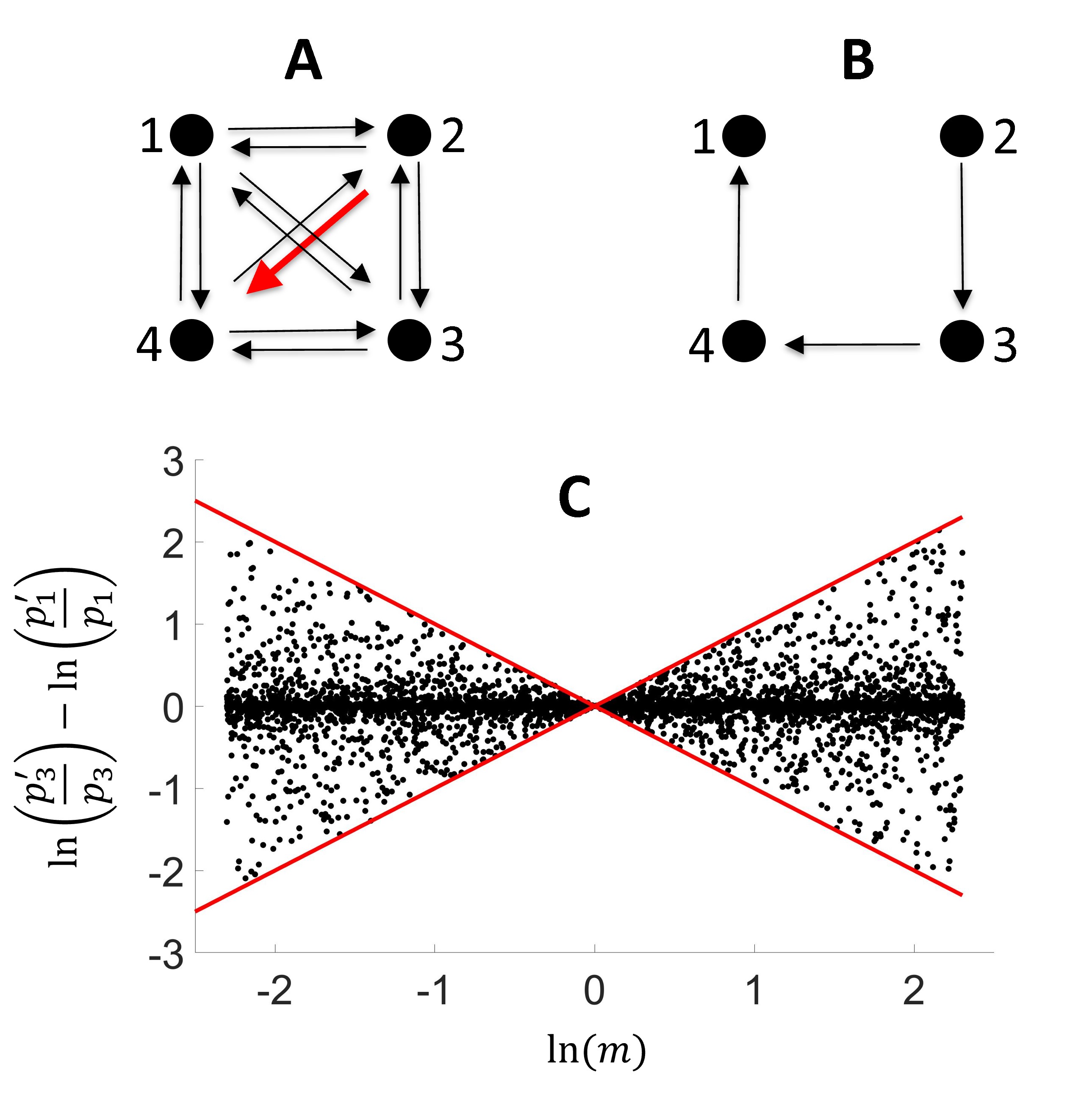}
\caption{\textbf{A} A linear framework graph representing a four-state Markov process, with the edge labels omitted for clarity. \textbf{B} A spanning tree of the graph in \textbf{A} with the root at vertex 1. \textbf{C} The black dots show $10^4$ data points obtained as described in the text, for which the expression in the fundamental inequality in Eq.\ref{fb2} is plotted against $\ln(m)$. The red lines show the bounds in Eq.\ref{fb2}.}
\label{Fig1}
\end{figure}

The linear framework is based on finite, directed graphs with labelled edges \citep{gunawardena2012linear,kmg22} (Fig.\ref{Fig1}A). Graph vertices, denoted $1, 2, \cdots, N$, represent mesostates of a system; edges, denoted $i \ra j$, represent transitions; and edge labels, denoted $\ell(i \ra j)$, represent positive transition rates, with dimensions of (time)$^{-1}$. Such a graph corresponds \citep{mirzaev2013laplacian} to a finite-state, continuous-time, time-homogeneous Markov process, $X(t)$, defined by the conditional distribution, $\Pr(X(t+h) = j | X(t) = i)$ for $h \geq 0$, in which the edge label is the infinitesimal transition rate, 
\begin{equation}\label{Ratedef}
    \ell(i \to j) = \lim_{\Delta t \to 0}\frac{\Pr(X(t+\Delta t) = j \mid X(t) = i)}{\Delta t} \,,
\end{equation}
whenever that rate is positive. We will treat graphs and Markov processes interchangeably in what follows. 

A linear framework graph specifies a dynamics on the probabilities of each vertex, $p_i(t) = \Pr(X(t) = i)$. This dynamics is most easily described by treating each edge as a chemical reaction under mass-action kinetics, with the edge label as the rate constant. Since an edge has only one source vertex, the resulting dynamics is linear and is therefore described by a matrix equation, 
\begin{equation}\label{Master}
    \frac{dp(t)}{dt} = {\cal L}(G) \cdot p(t) \,
\end{equation}
where ${\cal L}(G)$ is the Laplacian matrix of the graph. The entries of the Laplacian are given by, 
\begin{equation}\label{Lmatrix}
    {\cal L}(G)_{ji} = \begin{cases}
        \ell(i \to j) & \text{if $i \neq j$} \\
        -\sum_{k \neq i}{\ell(i \to k)} & \text{if $i=j$.}
    \end{cases}
\end{equation}
Since matter is neither created nor destroyed in this chemical interpretation, there is a conservation law, which corresponds to $1^{T}\cdot {\cal L}(G) = 0$. Eq.\ref{Master} is the master equation, or Kolmogorov forward equation, of the Markov process associated to $G$. The linear framework acquires its name from the linearity of the master equation. 

Let us reserve $p$ for the steady-state solution of Eq.\ref{Master}, as opposed to $p(t)$, which represents the time-dependent probabilities. That is, $p=\lim_{t \rightarrow \infty}p(t)$. The importance of this graph-theoretic approach is that the steady-state probabilities of $G$, $p$, can be expressed as rational functions of the labels by exploiting the Matrix-Tree theorem (MTT) of graph theory \citep{kmg22}. Because of Eq.\ref{Master}, $p$ lies in the kernel of the Laplacian matrix, $p \in \ker\lap(G)$, and is a right eigenvector for the zero eigenvalue of ${\cal L}(G)$. There is therefore no difficulty with expressing $p$ in terms of the principal minors of ${\cal L}(G)$. However, these minors are determinants with alternating signs. The MTT shows that remarkable cancellations take place in the minors of Laplacian matrices which yield a \emph{manifestly positive} rational expression for $p$. 

Provided $G$ is \emph{strongly connected}, so that any two distinct vertices are connected by a directed path of edges, the kernel of the Laplacian matrix is 1-dimensional, $\dim\ker\lap(G) = 1$, and the MTT yields a canonical basis element $\rho(G) \in \ker\lap(G)$ defined as follows. If $H$ is any subgraph of $G$, let $q(H)$ be the product of the edge labels over $H$, $q(H) = \prod_{i \ra j \in H} \ell(i \ra j)$. A spanning tree, $T$, is a subgraph of $G$ that includes every vertex (spanning) and has no cycles if edge directions are ignored (tree); it is rooted at $i$ if $i$ is the only vertex with no outgoing edges (Fig.\ref{Fig1}B). Let $\Theta_i(G)$ denote the set of spanning trees of $G$ rooted at $i$. A strongly connected graph has a spanning tree rooted at each vertex, so that $\Theta_i(G) \not= \emptyset$. Then, 
\begin{equation} 
\rho_i(G) = \sum_{T \in \Theta_i(G)} q(T) \,.
\label{e-mtt}
\end{equation}
This expression is a non-empty sum of monomials in the edge labels with coefficients $+1$ and is therefore manifestly positive. Since both $p$ and $\rho(G)$ are in $\ker\lap(G)$, 
\begin{equation}
p \propto \rho(G) \,,
\label{e-propto}
\end{equation}
by which we mean that $p_i = \lambda \rho_i(G)$, for some scalar $\lambda$. This proportionality constant may be removed by normalising to the total probability, so that
\begin{equation}
p_i = \frac{\rho_i(G)}{\rho_1(G) + \cdots + \rho_N(G)} \,,
\label{e-prob}
\end{equation}
where $N$ is the number of vertices. Eq.\ref{e-prob} expresses $p_i$ as a manifestly positive rational function of the edge labels. Such positivity is exactly what would be expected for the dependence of $p$ on the transition rates.

\subsection*{Path entropies}

We introduce a thermodynamic interpretation in the linear framework as follows. We set $k_B=T=1$, and work with dimensionless entropy and energy. Here, $k_B$ is the Boltzmann constant and $T$ is the ambient temperature.

We will assume from now on that each graph is \emph{reversible}, so that if $i \ra j$, then also $j \ra i$, where the reverse edge corresponds to the reverse of the process in the forward edge. Under \emph{local detailed balance} \citep{schnakenberg1976network,hill1966studies,seifert2008stochastic}, the total entropy change from $i$ to $j$, $S(i \lra j)$, which is the entropy change in the external reservoirs together with the internal entropy difference between $j$ and $i$, may be identified with the log label ratio 
\[ S( i \lra j) = \ln\left[\frac{\ell(i\ra j)}{\ell(j \ra i)}\right] \,.\]
Let $R(i,j)$ denote the set of directed paths from $i$ to $j$, $i = i_1 \rightarrow i_2 \rightarrow \cdots \rightarrow i_k = j$. If all the vertices along a path are distinct, then the path is called \emph{minimal}; the set of minimal directed paths from $i$ to $j$ will be denoted $M(i,j) \subseteq R(i,j)$. $R(i,j)$ is infinite but $M(i,j)$ is always finite. If $P$ is a directed path, let $P^{\circ}$ denote the corresponding path of reversible edges, $i = i_i \lra i_2 \lra \cdots \lra i_k = j$. If $P \in R(i,j)$, then the total entropy change along $P$, $S(P)$, is just the sum of the entropy changes along each pair of reversible edges in $P^{\circ}$,  
\begin{equation}
S(P) = \sum_{u = 1}^{k-1} S(i_u \lra i_{u+1}) \,.
\label{pathentropy}
\end{equation}

We refer to Eq.\ref{pathentropy} as path entropy. If the path $C$ is a cycle, so that $i = i_1 = i_k = j$, then $S(C)$ is called the \emph{thermodynamic affinity} of $C$ and denoted $A(C)$ \citep{schnakenberg1976network}. A graph (or the associated Markov process) can reach thermodynamic equilibrium if, and only if, $A(C) = 0$ for each cycle $C$. This \emph{cycle condition} is equivalent to detailed balance or microscopic reversibility \citep{kmg22}. 

If the cycle condition holds, Eq.\ref{Master} relaxes to a steady-state probability distribution at thermodynamic equilibrium. We will generically denote such distributions by $p^{eq}$. Given an arbitrary reference vertex, $1$, choose any minimal path $P_i \in M(i,1)$. It is not difficult to see \citep{kmg22}, that when the cycle condition holds, an alternative basis element may be constructed in the Laplacian kernel, $\mu(G) \in \ker\lap(G)$, where,
\begin{equation}
\mu_i(G) =\mathrm{e}^{-S(P_i)} \,.
\label{e-mu}
\end{equation}
The independence of $S(P_i)$ from the choice of minimal path $P_i$ is a simple consequence of the cycle condition. Note that if another reference vertex, $q$, is chosen and and a path $P'_i \in M(i,q)$ is chosen, then, again because of the cycle condition, $S(P'_i) = S(P_i) + S(Q)$, where $Q$ is any path in $R(1,q)$. It follows from Eq.\ref{e-mu} that $\mu(G)$ changes by the scalar multiple $e^{-S(Q)}$ when the reference vertex changes from $1$ to $q$.  

Because $\mu(G)$ is a basis element for $\ker\lap(G)$, 
\begin{equation}
 p^{eq}_i \propto \mathrm{e}^{-S(P_i)} \,,
\label{eql}
\end{equation}
so that $p^{eq}_i$ is given by a similar rational expression to that in Eq.\ref{e-prob} but with $\mu(G)$ in place of $\rho(G)$. (The scalar multiple $e^{-S(Q)}$ disappears, as required to make $p^{eq}$ independent of the choice of reference vertex.) As mentioned previously, the proportionality constant can be determined from the normalization condition, $p_1^{eq} + \cdots + p^{eq}_N= 1$, which gives,
\begin{equation}
p^{eq}_i = \frac{\mathrm{e}^{-S(P_i)}}{\sum_{j} \mathrm{e}^{-S(P_j)}} \,.
\label{normalization}
\end{equation}
This rational expression for $p^{eq}_i$ is the usual prescription of equilibrium statistical mechanics, with the denominator being the partition function. Importantly, Eq.\ref{e-prob} continues to hold away from thermodynamic equilibrium, so the graph-theory of the linear framework offers a setting in which non-equilibrium statistical mechanics is exactly solvable in terms of rational functions of the transition rates. This can be very useful mathematically \citep{wcg18,kmg22} but the combinatorics of spanning trees rapidly becomes intractable, even for relatively simple graphs, and the resulting algebra lacks the thermodynamic meaning of Eq.\ref{eql} at equilibrium. The next section explains how Eq.\ref{eql} can be smoothly generalised away from equilibrium. 

\subsection*{The arboreal distribution}

When the cycle condition does not hold, $S(P_i)$ depends on the choice of $P_i$ and Eq.\ref{eql} is no longer valid. To address this, we define a probability distribution on the set of spanning trees rooted at the reference vertex, $\Theta_1(G)$, in which each tree $T$ acquires the probability \citep{ccetiner2022reformulating}, 
\begin{equation}\label{arboreal}
\pr_{\Theta_1(G)}(T) = \frac{q(T)}{\sum_{T \in \Theta_1(G)} q(T)} \,.
\end{equation}
Given any $T \in \Theta_1(G)$, there is a unique minimal path, denoted $T_i \in M(i,1)$, from $i$ to $1$. Then an alternative basis element, which was originally derived in \citep{ccetiner2022reformulating}, may be given as,
\begin{equation}
\tilde{\rho}_i(G) = \sum_{T \in \Theta_1(G)} \pr_{\Theta_1(G)}(T)\mathrm{e}^{-S(T_i)}=\langle \mathrm{e}^{-S(T_i)}\rangle \,,
\label{rhoness}
\end{equation}
where the angular brackets denote an average taken over the arboreal distribution \citep{ccetiner2022reformulating}. It follows that, if $p^{neq}$ denotes a non-equilibrium steady-state probability distribution, this is proportional to an average over the arboreal distribution on $\Theta_1(G)$ \citep{ccetiner2022reformulating}, 
\begin{equation}\label{NessFinal1}
p_i^{neq} \propto \tilde{\rho}_i(G)= \langle \mathrm{e}^{-S(T_i)} \rangle\,.
\end{equation}
Similar to Eq.\ref{normalization}, the steady-state probabilities are recovered using the normalization condition: 
\begin{equation}\label{NessFinal}
p_i^{neq}=\frac{\langle \mathrm{e}^{-S(T_i)}\rangle}{\sum_j \langle \mathrm{e}^{-S(T_j)}\rangle} \,.
\end{equation}.

If the cycle condition holds, then $S(T_i)$ does not depend on $T$, the average is just the unique entropy of minimal paths in $M(i,1)$ and Eq.\ref{NessFinal} smoothly reduces to Eq.\ref{normalization}. At equilibrium, the steady-state probability is proportional to the exponential of the minimal path entropy (Eq.\ref{eql}); away from equilibrium it is proportional to the average, over the arboreal distribution, of the exponential of the minimal path entropies (Eq.\ref{NessFinal1}). Note that $\tilde{\rho}_i(G)$ also gives a manifestly positive rational function of the edge labels for $p_i$. In addition, this reformulation tames the algebraic complexity because the number of distinct path entropies depends only on the number of edges where energy is expended, not on the size of the system. For example, when energy is dissipated at a single transition, there are only 3 distinct minimal path entropies, no matter how large the system is \citep{ccetiner2022reformulating}. Describing non-equilibrium steady states through a nonlinear average of path entropies has been a valuable approach in the study of non-equilibrium systems. With this background, we can proceed to our main findings. 

\section*{Results}

\subsection*{Fundamental bound for a single energetic edge}

Let $G$ be a graph, which need not satisfy the cycle condition, and let $p$ be the corresponding steady-state probability distribution. Choose any edge, $z_1 \ra z_2$, and consider the new graph $G'$ in which the label on this edge is multiplied by $m > 0$: $\ell(z_1 \ra_{G'} z_2) = m\ell(z_1 \ra_{G} z_2)$. We will refer to $z_1 \ra z_2$ as the \emph{energetic edge}. The corresponding Markov process relaxes to a new steady state by construction, if $m \neq 1$. If the underlying system is at thermal equilibrium, the system now relaxes to a non-equilibrium steady state. However, in its most general
form, the system can be viewed as transitioning from an initial arbitrary non-equilibrium steady state to another as a result of the perturbation. Let $p'$ be the resulting steady-state probability distribution of $G'$. Let $i$ and $j$ be any two vertices in $G$ and $G'$. Then, 
\begin{equation} \label{fb2}
 -|\ln(m)|\leq\ln\left(\frac{p'_i}{p_i}\right) - \ln\left(\frac{p'_j}{p_{j}}\right) \leq |\ln(m)| \,.  
\end{equation}

The proof of this fundamental inequality is as follows. If we choose $z_1$ as the reference vertex in $G$ and in $G'$ then Eq.\ref{rhoness} tells us that,
\begin{equation}
\tilde{\rho}_i(G) = \sum_{T \in \Theta_{z_1}(G)} \pr_{\Theta_{z_1}(G)}(T)\mathrm{e}^{-S(T_i)}
\label{e-rgip1}
\end{equation}
and similarly that, 
\begin{equation}
\tilde{\rho}_i(G') = \sum_{T' \in \Theta_{z_1}(G')} \pr_{\Theta_{z_1}(G')}(T')\mathrm{e}^{-S(T'_i)} \,.
\label{e-rgip2}
\end{equation}
Note that $G$ and $G'$ have identical vertices and edges and differ only in the labels on the energetic edge. Because the energetic edge leaves $z_1$, it cannot be an edge in any $T \in \Theta_{z_1}(G)$. It follows that $\Theta_{z_1}(G) = \Theta_{z_2}(G)$. If we let $T' \in \Theta_{z_1}(G')$ be the tree corresponding to $T \in \Theta_{z_1}(G)$, then, since $m$ does not appear as a label on these spanning trees, $q(T) = q(T')$. Accordingly, the arboreal distribution is the same for both graphs. However, the energetic edge may appear on the path of reversible edges $T'^{\circ}_i$ and therefore also in $\mathrm{e}^{-S(T'_i)}$. It follows from Eq.\ref{pathentropy} that $\mathrm{e}^{-S(T'_i)}$ in $G'$ is given by, 
\begin{equation}
\begin{array}{ll}
  \mathrm{e}^{-S(T_i)} & \text{if $z_1 \ra z_2 \notin T_i'^{\circ}$} \\
  m\mathrm{e}^{-S(T_i)} & \text{if $z_1 \ra z_2 \in T_i'^{\circ}$}\,.
\end{array}
\label{e-xps}
\end{equation}
With this in mind, let us partition $\Theta_{z_1}(G)$ into disjoint subsets as follows,
\begin{align*}
\Theta_{z_1,0}(G) & = \{ T \in \Theta_{z_1}(G) \,|\, z_1 \ra z_2 \not\in T_i^{\circ} \} \\
\Theta_{z_1,1}(G) & = \{ T \in \Theta_{z_1}(G) \,|\, z_1 \ra z_2 \in T_i^{\circ} \} \,,
\end{align*}
and define the quantities $A$ and $B$ by, 
\begin{align*}
A & = \sum_{T \in \Theta_{z_1,0}(G)} \pr_{\Theta_{z_1}(G)}(T)\mathrm{e}^{-S(T_i)} \\
B & = \sum_{T \in \Theta_{z_1,1}(G)} \pr_{\Theta_{z_1}(G)}(T)\mathrm{e}^{-S(T_i)} \,.
\label{e-rgip}
\end{align*} 
Since $z_2 \ra z_1$ is a minimal path in $M(z_2, z_1)$, there will be spanning trees, $T \in \Theta_{z_1}(G)$, that contain this edge, so that the reverse energetic edge $z_1 \ra z_2$ is in $T_i^{\circ}$. It follows that $B \not= 0$. It is possible that there is no other minimal path in $M(z_2,z_1)$, so that $A = 0$, but this can happen only if $z_1 \lra z_2$ is a \emph{bridge} that links two disjoint subgraphs of $G$. Accordingly, 
\[ A \geq 0 \hspace{1em}\text{and}\hspace{1em} B > 0 \,.\]

Using Eqs.\ref{e-rgip1} and \ref{e-rgip2}, the fact that $\Theta_{z_1}(G) = \Theta_{z_1}(G')$ and Eq.\ref{e-xps}, we see that,
\begin{align*}
\tilde{\rho}_i(G) & = A + B \\
\tilde{\rho}_i(G') & = A + mB \,,
\end{align*}
so that, 
\begin{equation}
\frac{\tilde{\rho}_i(G')}{\tilde{\rho}_i(G)} = \frac{A + mB}{A + B} \,.
\label{e-rig}
\end{equation}
If $m \geq 1$, then $0 \leq A \leq mA$ and $B \leq mB$, so that $A + B \leq A + mB \leq m(A + B)$. Since $0 < A + B$, we see that $1 \leq (A + mB)/(A + B) \leq m$. Hence, by Eq.\ref{e-rig},
\begin{equation}
1 \leq \frac{\tilde{\rho}_i(G')}{\tilde{\rho}_i(G)} \leq m \,.
\label{e-frh1}
\end{equation}
By construction, $\tilde{\rho}_{z_1}(G) = \tilde{\rho}_{z_1}(G')=1$. We can therefore rewrite Eq.\ref{e-frh1} as,
\[
1 \leq \left(\frac{\tilde{\rho}_i(G')}{\tilde{\rho}_{z_1}(G')}\right) \left/ \left(\frac{\tilde{\rho}_i(G)}{\tilde{\rho}_{z_1}(G)}\right) \right. \leq m \,.
\]
But now, we can make use of Eq.\ref{NessFinal1}, to replace $\tilde{\rho}$ by the steady-state probabilities, $p$ for $G$ and $p'$ for $G'$, so that,
\[
1 \leq \left(\frac{p'_i}{p'_{z_1}}\right) \left/ \left(\frac{p_i}{p_{z_1}}\right)\right. \leq m \,.
\]
Rearranging and applying $\ln$, we see that,
\begin{equation}
0 \leq \ln\left(\frac{p'_i}{p_i}\right) - \ln\left(\frac{p'_{z_1}}{p_{z_1}}\right) \leq \ln(m) \,.    
\label{e-frh2}
\end{equation}
Since Eq.\ref{e-frh2} holds for any vertex $i$, it is not difficult to see that it implies that, given any other vertex $j$,
\[
 -\ln(m)\leq\ln\left(\frac{p'_i}{p_i}\right) - \ln\left(\frac{p'_j}{p_{j}}\right) \leq \ln(m) \,.    
\]
We leave it to the reader to supply the corresponding argument for $m \leq 1$, to finally conclude that, no matter what the value of $m$, 
\[
 -|\ln(m)|\leq\ln\left(\frac{p'_i}{p_i}\right) - \ln\left(\frac{p'_j}{p_{j}}\right) \leq |\ln(m)| \,,  
\]
which yields Eq.\ref{fb2}. This completes the proof.

We can give a thermodynamic interpretation to Eq.\ref{fb2} as follows. Let $C$ be any cycle of reversible edges in $G$ that contains the energetic edge and let $C'$ be the corresponding cycle in $G'$. It follows from Eq.\ref{pathentropy} that the change in affinities of these cycles is given by $\Delta A(C) = A(C') - A(C) = \pm \ln(m)$, where the sign depends on whether the energetic edge points in the forward or reverse direction around the cycle. We can therefore rewrite Eq.\ref{fb2} as
\begin{equation}
 -|\Delta A(C)| \leq\ln\left(\frac{p'_i}{p_i}\right) - \ln\left(\frac{p'_j}{p_{j}}\right) \leq |\Delta A(C)|  \,,  
\label{boundsfinal}
\end{equation} 
where $C$ is any cycle of reversible edges that contains the energetic edge, $z_1 \ra z_2$. 

Eqns.\ref{fb2} and \ref{boundsfinal} are universally valid for any Markov process specified by a reversible, strongly-connected linear framework graph, regardless of the size or topology of the graph, the location of the energetic edge and the values of the edge labels (transition rates).

The question arises as to whether the bounds in Eqns.\ref{fb2} and \ref{boundsfinal} are tight. Fig.\ref{Fig1}C shows the results of a numerical simulation for the graph in Fig.\ref{Fig1}A. We sampled transition rates as $10^a$, where $a$ was drawn independently at random from the uniform distribution on $(-3,3)$. After the corresponding steady-state, $p$, is obtained, the label on the energetic edge, $2\ra 4$, shown in red is altered by multiplying it by $m$, to yield $m\ell(2 \ra 4)$, and the new steady-state solution $p'$ is calculated  Here, $m$ is chosen randomly from the uniform distribution on $(0,10)$. Since the parameters are sampled randomly, both $p$ and $p^\prime$ are non-equilibrium steady states. We see from Fig.\ref{Fig1}C that the bound in Eq.\ref{fb2} appears tight. Following this same procedure, we tested the tightness for the three different graphs shown in Fig.\ref{Fig2} and found each example to also be tight. Note that the fundamental bound is universal as mentioned earlier but it is possible to exploit the topology of energy expenditure and improve the fundamental bound further. This is already hinted in Eq.\ref{e-frh2}, which shows that if one of the vertices is $z_1$, which is the source vertex of the energetic edge, the lower bound becomes zero instead of $-\ln(m)$ for $m>1$. We show an instance of this in Fig.\ref{Fig2}C.


\subsection*{Bound on the Kullback-Leibler divergence}

The Kullback-Leibler (KL) divergence between two finite probability distributions, $p'$ and $p$, on $N$ states, is defined as \citep{kullback1951information,kullback1997information,cover1999elements},
\begin{equation}
\dkl(p^{\prime}||p) \equiv \sum_{i=1}^N p^{\prime}_i \ln\left(\frac{p^{\prime}_i}{ p_i}\right) \,.
\end{equation}
KL divergence is an asymmetric distance measure for probability distributions and is discussed further below. Our second main result is a bound on $\dkl$ for the same situation considered in the previous section, with a single energetic edge. Using the same notation as previously, $p$ is the steady-state probability distribution of $G$ and $p'$ is the steady-state distribution of $G'$, which is obtained by altering the label on the energetic edge by a factor of $m$. Then, 
\begin{equation}
\text{D\textsubscript{KL}}(p^{\prime}||p) \leq |\ln(m)| = |\Delta A(C)| \,,
\label{KLD}
\end{equation}
where $C$ is any cycle in $G$ that contains the energetic edge. The last equality simply restates what was shown in the previous section. 

Eq.\ref{KLD} follows readily from the fundamental inequality. We may assume that $p' \not= p$, for otherwise Eq.\ref{KLD} is obvious. Let $N$ be the number of vertices in $G$ and $G'$. Define $K_i = p'_i - p_i$. Since $\sum_{i = 1}^N p'_i = \sum_{i=1}^N p_i = 1$, we have $\sum_{i = 1}^N K_i = 0$. Since $p' \not= p$, there must be some indices for which $K_i > 0$ and some for which $K_i < 0$. Accordingly, there are some indices for which $p'_i/p_i = 1 + K_i/p_i > 1$ and some for which $p'_i/p_i < 1$. Let $u$ be an index for which 
\[ \frac{p'_u}{p_u} = \max_{1 \leq i \leq N}\frac{p'_i}{p_i} \,,\] 
so that $p'_u/p_u > 1$ and let $v$ be any index for which $p'_v/p_v < 1$. Then, $\ln(p'_u/p_u) > 0$ and $\ln(p'_v/p_v) < 0$. Because the average is always less than the maximum, we have, by Eq.\ref{fb2},
\[
\begin{split}
\dkl(p'||p) & \leq \ln(p'_u/p_u) < \ln(p'_u/p_u) - \ln(p'_v/p_v) \\
            & \leq |\ln(m)| = |A(C)| \,.
\end{split}
\]
which gives Eq.\ref{KLD}. This completes the proof.

\subsection*{Multiple Energetic Edges}

If multiple edge labels are altered, it is easy to extrapolate the fundamental inequality. However, we note that this ignores the interactions between the changes, as explained further in the Discussion. We only point out the argument for two energetic edges, since the general case is evident. Let $z_1\rightarrow z_2$ and $z_3\rightarrow z_4$ be the arbitrarily chosen energetic edges such that $\ell(z_1 \ra_{G'} z_2) = m_1\ell(z_1 \ra_{G} z_2)$ and $\ell(z_3 \ra_{G'} z_4) = m_2\ell(z_3 \ra_{G} z_4)$, with $m_1>0$ and $m_2>0$. Let us denote the corresponding unique steady-state distribution by $p''$. We can think of the case with two energetic edges as two separate single-energetic-edge perturbations as follows: If we were to first alter a single edge label, say, $\ell(z_1 \ra z_2)$, as above, wait for the system to reach steady state ($p^\prime$), and then change the second edge label, $\ell(z_3 \ra z_4)$, the resultant steady-state distribution ($p''$) would be the same distribution if these two edge labels were altered simultaneously. Hence, the following chain of inequalities can be utilized to connect $p$ to $p''$ as follows,
\begin{equation}
\begin{split}
-|\ln(m_1)|\leq \ln\left(  \frac{p^\prime_i}{p^\prime_j} \right) -\ln\left( \frac{p_i}{p_{j}}\right) \leq |\ln(m_1)|\\
-|\ln(m_2)|\leq \ln\left(  \frac{p''_i}{p''_{j}} \right) -\ln\left( \frac{p'_i}{p'_{j}}\right) \leq |\ln(m_2)|.
\end{split}
\end{equation}
Summing up these two inequalities, we obtain the inequality below, and establish a connection between $p$ and $p''$ through $p\rightarrow p' \rightarrow p''$, 
\begin{equation}
\begin{split}
-|\ln(m_1)|-|\ln(m_2)|&\leq \ln\left(  \frac{p''_i}{p''_j} \right) -\ln\left( \frac{p_i}{p_{j}}\right)\\
& \leq  |\ln(m_1)|+|\ln(m_2)|.
\end{split}
\end{equation}
When there are $n$ (different) energetic edges, using the same steps as above, we can establish a similar connection between the final non-equilibrium steady-state, denoted by $p^{(n)}$, and the initial distribution $p$: 
\begin{equation}\label{Uniboundmultiple}
\begin{split}
- \sum_{i=1} ^n |\ln(m_i)|&\leq  \ln\left(  \frac{p^{(n)}_i}{p^{(n)}_j} \right) -\ln\left( \frac{p_i}{p_{j}}\right)\\ 
&\leq \sum_{i=1} ^n |\ln(m_i)|.
\end{split}
\end{equation}

\section*{Discussion}

The KL divergence often makes an appearance at the interface of non-equilibrium physics and information theory, where it has been shown to be related to familiar thermodynamic quantities \citep{schlogl1971stability,gaveau1996master,kawai2007dissipation,vaikuntanathan2009dissipation}. Of particular relevance here is the work of \cite{gaveau1996master}. Consider a system at thermodynamic equilibrium, where the states of the system are populated according to the Boltzmann distribution: if state $i$ has energy $E_i$, then the equilibrium probability of $i$ is given by $p^{eq}_i=\mathrm{e}^{-E_i}/Z$, where $Z$ is the partition function. In this case, the equilibrium free energy of the system is defined as $F^{eq}=\langle E \rangle_{p^{eq}}-{\cal S}(p^{eq})$, where $\langle E \rangle_{p^{eq}}$ is the average energy, $\langle E \rangle_{p^{eq}} = \sum_i p_i^{eq} E_i$, and ${\cal S}(p^{eq})$ is the Gibbs-Shannon entropy of $p^{eq}$ given by ${\cal S}(p^{eq})=-\sum_i p_i^{eq}\ln(p_i^{eq})$. By analogy to the equilibrium setting, the non-equilibrium free energy of a non-equilibrium state, $p^{neq}$, may be defined as $F^{neq}=\langle E \rangle_{p^{neq}} - {\cal S}(p^{neq})$ \citep{gaveau1996master,parrondo2015thermodynamics}. The difference between these two free energies is then exactly the KL divergence between $p^{neq}$ and $p^{eq}$ \citep{gaveau1996master},
\begin{equation}
\dkl(p^{neq}||p^{eq}) = F^{neq}-F^{eq} \,.
\label{difffree}
\end{equation}
Schl\"{o}g has named the free-energy difference, $F^{neq}-F^{eq}$, the \emph{availability}; it corresponds to the maximum available work that can be produced by a system whose equilibrium and non-equilibrium distributions are given by $p^{eq}$ and $p^{neq}$, respectively,  \citep{schlogl1988thermodynamic} (see also \cite[Box 1]{parrondo2015thermodynamics}). Combining Eq.\ref{difffree} with Eq.\ref{KLD}, we see that, when $p^{neq}$ is obtained from $p^{eq}$ by perturbing a single edge-label by a factor of $m$ then, 
\begin{equation}\label{DeltaF}
F^{neq} - F^{eq}\leq |\ln(m)| = |\Delta A(C)| \,,
\end{equation}
where $C$ is any cycle that contains the energetic edge. 

Unlike the fundamental bound, Eq.\ref{KLD} is not tight as shown in Fig.\ref{Fig2KLD}. It is an interesting future direction to determine a tighter bound on the KL divergence. Another important point is to understand how to exploit the topology of energy expenditure to maximize the free energy difference between equilibrium and non-equilibrium systems at a fixed energy budged. Living systems might have already mastered this as they constantly carry out information processing both at equilibrium and away from equilibrium and especially they have devised strategies to navigate the constraints imposed by equilibrium thermodynamics as in the case of kinetic proofreading. Analyzing the working principles of dissipative biological systems can provide some useful insights into this problem. A final point worth noting is that when there are multiple energetic edges, their perturbations can interact and intricately influence a system's information processing capacity. This complexity is particularly relevant when these energetic edges exert opposing effects, as in the case when one parameter ($m_1$) is greater than 1 and another ($m_2$) is less than 1. It would be interesting to investigate what kind of new capabilities can be gained through these interference patterns. All these result are derived by expressing non-equilibrium steady probabilities in terms of minimal path entropies. With this new way of looking at non-equilibrium steady states, we can now ``follow the energy" to understand its functional significance in information processing. 

\bibliography{Universalbounds}
\vspace*{1cm}
\appendix
\beginsupplement
\section*{Supplementary Information}
\begin{figure}[h]
\includegraphics[width=0.45\textwidth]{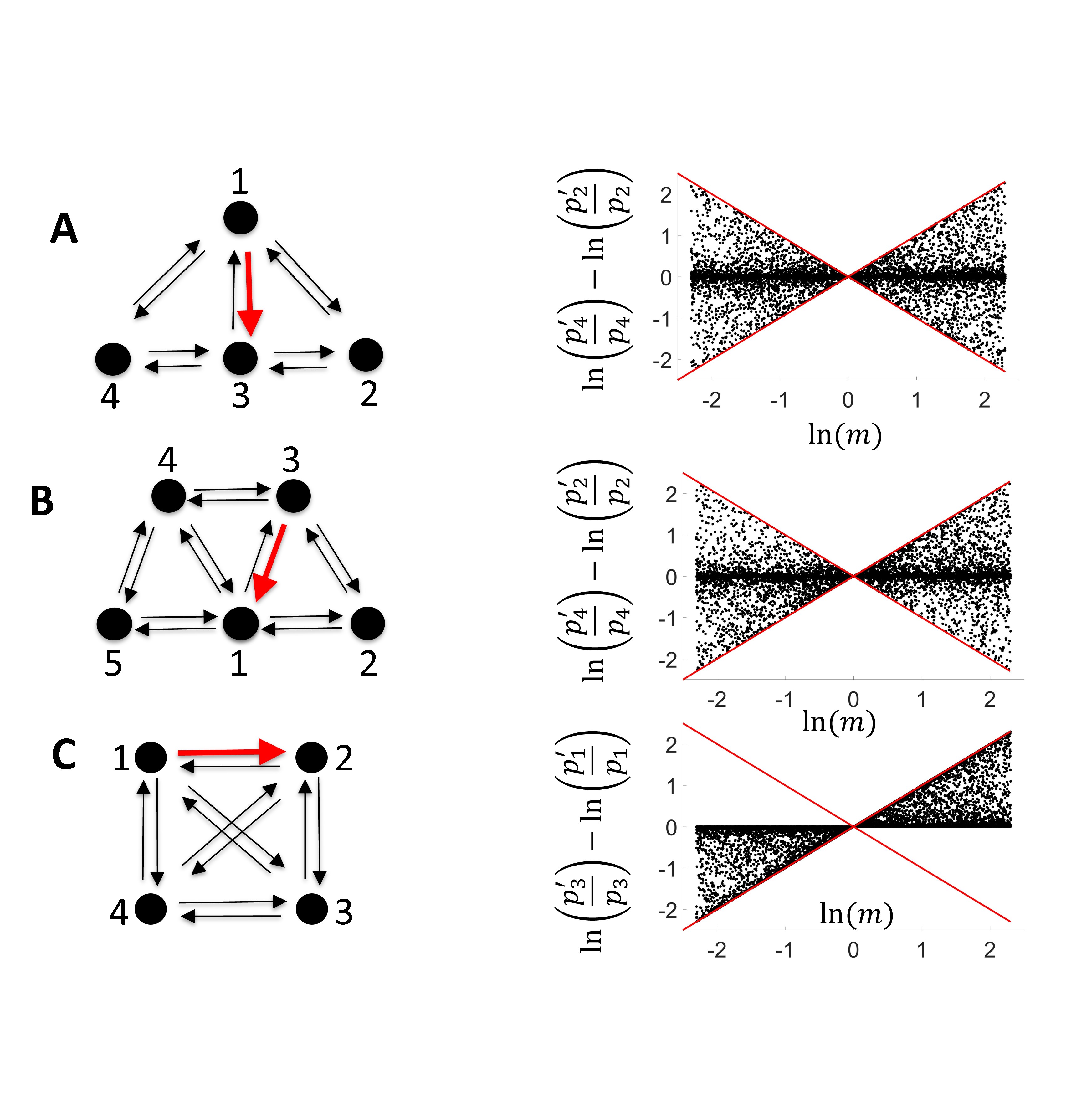}
\caption{Testing the tightness of the fundamental bound (Eq.\ref{fb2}) for a single energetic edge. \textbf{A, B, C} show on the left three example graphs with the energetic edge in red. The corresponding plot on the right shows the expression in the fundamental inequality in Eq.\ref{fb2} plotted against $\ln(m)$, for randomly sampled parameter sets obtained as described in the text. The red lines show the bound in Eq.\ref{fb2}.}
\label{Fig2}
\end{figure}
\begin{figure}
\includegraphics[width=0.45\textwidth]{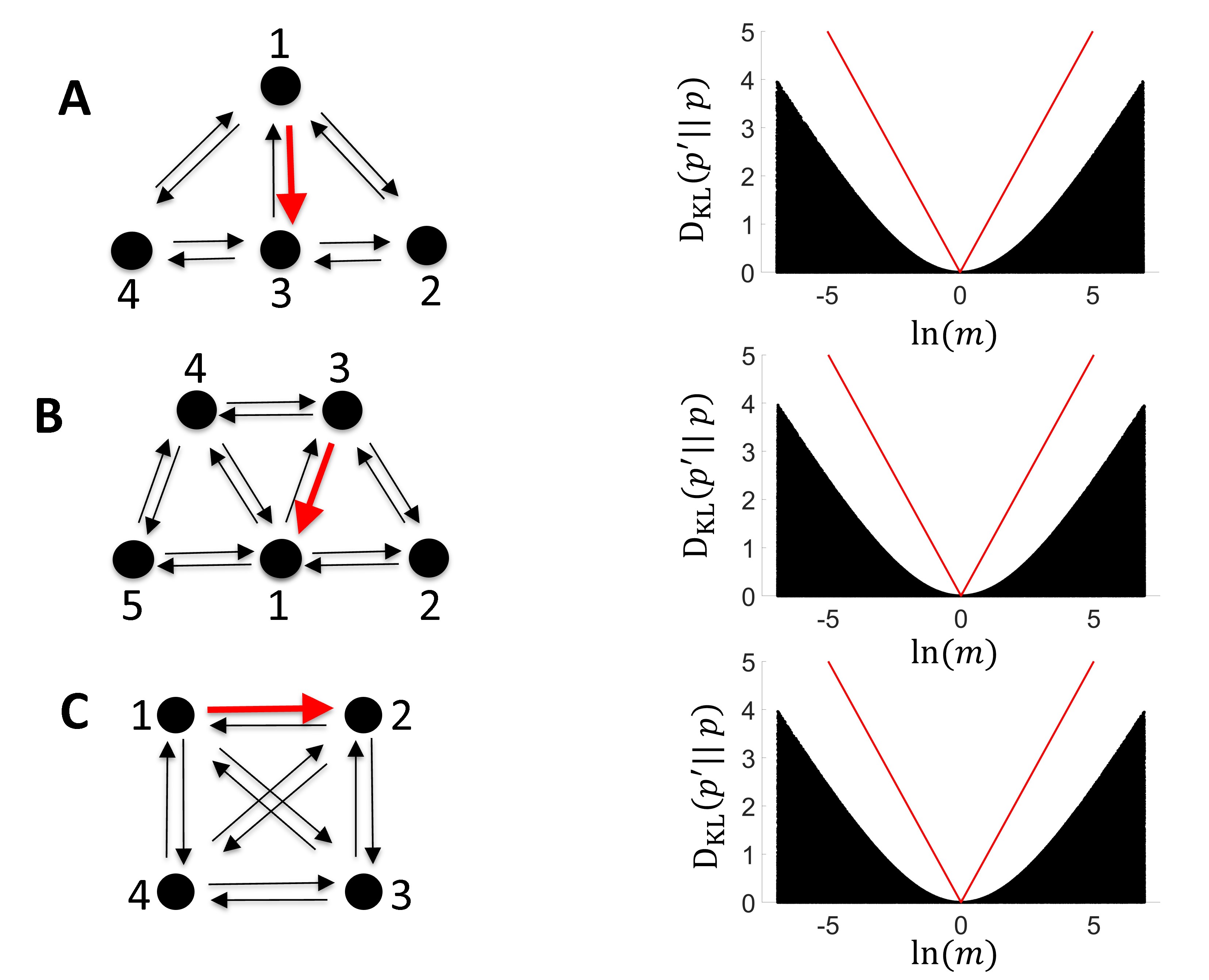}
\caption{The Kullback-Leibler (KL) divergence in Eq.\ref{KLD} in relation to the fundamental inequality in Eq.\ref{fb2}. \textbf{A, B, C} show on the left the same systems as in Fig.\ref{Fig2}. The corresponding plots on the right show the KL divergence from Eq.\ref{KLD} against $\ln(m)$, where $m$ is the multiple on the energetic edge (red), for $10^6$ data points, randomly sampled as in Fig.\ref{Fig2}. The red lines in the plots show the bounds in Eq.\ref{fb2}. From these numerical results, we see the existence of a tighter bound on the KL divergence.}
\label{Fig2KLD}
\end{figure}
\end{document}